\documentclass[10pt]{article}
\usepackage[dvipsnames]{xcolor}

\usepackage{inputenc, graphicx, url}

\usepackage{fullpage}

\usepackage{amsmath,amsthm,amssymb}

 \usepackage{algorithm}
 \usepackage{algpseudocode}
 
\date{}

\newtheorem{thm}{Theorem}
\newtheorem{corollary}[thm]{Corollary}
\newtheorem{lem}[thm]{Lemma}
\newtheorem{remark}[thm]{Remark}

\newtheorem{proposition}[thm]{Proposition}
\newtheorem{definition}[thm]{Definition}

\newtheorem*{claim*}{Claim}

\title{On the Geometry of Stable Steiner Tree Instances}

\author{James Freitag\thanks{Department of Mathematics, Statistics, and Computer Science,
        University of Illinois at Chicago, {\tt jfreitag@uic.edu}. Supported in part by NSF CAREER award 1945251.}
        \and
        Neshat Mohammadi\thanks{Department of Computer Science,
        University of Illinois at Chicago, {\tt nmoham24@uic.edu}. Supported in part by NSF grant CCF-1934915.}
        \and
        Aditya Potukuchi\thanks{Department of Mathematics, Statistics, and Computer Science,
        University of Illinois at Chicago, {\tt adityap@uic.edu}. Supported in part by NSF grant CCF-1934915.}
        \and
        Lev Reyzin\thanks{Department of Mathematics, Statistics, and Computer Science,
        University of Illinois at Chicago, {\tt lreyzin@uic.edu}. Supported in part by NSF grant CCF-1934915.}
}

\begin{document}

\maketitle

\begin{abstract}

In this note we consider the Steiner tree problem
under Bilu-Linial stability.
We give strong geometric structural properties that need to be satisfied
by stable instances. 
We then make use of, and strengthen, these geometric properties to show 
that $1.562$-stable instances of Euclidean Steiner trees are polynomial-time
solvable. We also provide a connection between certain approximation
algorithms and Bilu-Linial stability for Steiner trees.    
    
\end{abstract}

\section{Introduction and previous work}

In this note, we initiate the study
of Steiner tree instances
that are stable to multiplicative perturbations
to the distances in the underlying
metric.
Our analysis lies in the Bilu-Linial stability~\cite{BiluL12} setting, 
which provides a way to study tractable instances of NP-hard problems.

Instances that are  $\gamma$-stable in the Bilu-Linial model have the property that the structure
of the optimal solution is not only unique, but also does not change even when the underlying
distances among the input points are perturbed by a multiplicative
factor $\gamma > 1$.  In their original paper, Bilu and Linial
analyzed MAX-CUT clustering, and since their seminal work, other
problems have been analyzed including center-based 
clustering~\cite{AwasthiBS12,BalcanL16,Ben-DavidR14},
multi-way cut problems~\cite{MakarychevMV14}, and metric TSP~\cite{MihalakSSW11}.\footnote{Bilu-Linial stability is
one among other notions of data stability studied in the literature~\cite{AckermanB09,BalcanBG09}. This is in
contrast to notions of of algorithmic stability,
which 
focus on properties algorithms as opposed to data, see e.g.~\cite{Alabdulmohsin15,Ben-DavidLP06,FanIMSSV20,LiuLNT17}.}

Here, we look at the metric Steiner tree problem%\footnote{The metric
%Steiner tree problem is known to be fully general, in the sense that an 
%arbitrary Steiner tree instance can be transformed to a metric instance
%that preserves approximation factor.} 
and also the more
restricted Euclidean version.  For general metrics, the Steiner tree 
problem is known to be APX-hard in the worst case~\cite{ChlebikC08}. For the Euclidean metric, a PTAS is known~\cite{Arora98}.

In this paper
we begin by providing strong geometric structural properties that need to be satisfied
by stable instances. These point to the existence of algorithms for non-trivial
families.  We then make use of, and strengthen, these geometric properties to show
that $1.562$-stable instances of Euclidean Steiner trees are polynomial-time
solvable. Finally, we discuss the connections between certain approximation
algorithms and Bilu-Linial stability for Steiner trees.

\section{Model and definitions}

%to form the foundation for understanding  $\gamma$-stable instances of Steiner tree that has been modeled in the next sections.For various values of $\gamma.$ These definitions are going to be used in other sections. We Start this section by defining various settings of Steiner trees problem and then we continue by defining what notion of stability has been studied in this work. Finally, we present the definition of $\gamma$-stable instances of Steiner tree. 

In this section, we recall the relevant definitions.
Fist we define the Steiner tree problem, which is
among Karp's 21 original NP-hard problems~\cite{Karp72}.
It has various applications including in network design, circuit layouts, and phylogenetic tree reconstruction.

\begin{definition}[{\bf the Steiner tree problem}]
For an undirected  graph
$G = (V, E)$ with edge weights $w_{e} \in \mathbb{R}^{+}_{0}$ for every edge $e \in E$, and a set $T \subseteq V$ of terminals. A Steiner tree $S$ is a tree in the graph $G$ that spans all terminal 
vertices $T$ and may contain some of the
non-terminals (also called Steiner points). 
The goal is to find such a tree of lowest weight,
which we call $\mathrm{OPT},$
$$\mathrm{OPT} = \arg\min_{S}\sum_{e\in S} w_e.$$
\end{definition}

%\begin{definition}[metric Steiner tree]
%Let $N$ be a set of points, $d$ be a
%metric $d: N \cross N \rightarrow \mathbb{R}^{+}_{0}$,
%and $T \subseteq N$ as set of of terminals. 
%...
%A Steiner tree is a tree in the graph $G$, which spans all terminals in $T$ and %may contain some of the
%Steiner points. The goal is to minimize the total weight $\sum
%e\in E(T) w(e)$ of the
%computed Steiner tree $T \subseteq G$. 
%\end{definition}

We can assume without loss of generality\footnote{For
any graph with distances specified on edges, a metric can be formed by taking the vertices to be points and considering the shortest path distances in the graph between pairs of points vertices.  Solving (or approximating) the Steiner tree problem on a metric formed in this matter solves (or approximates) the problem on the original graph.  See Vazirani~\cite{Vazirani01} for further discussion of this issue.}
that the vertices are points in a metric space
and the weights of the edges are given by the distance function -- when the input is in the form of a metric, we call this the
\textbf{metric Steiner tree problem}. Our results use properties of metric spaces, but move freely between the metric space and graph representations of the problem. When the metric is Euclidean,
this is called the \textbf{Euclidean Steiner tree problem}.

%Steiner tree where we have N points $(\left | V \right | = N)$ and a distance
%function $d: E \Rightarrow \mathbb{R}^{+}$on the edges.

%\begin{definition}[Geometric Steiner tree]
%Given N points in $\mathbb{R}^{d}$, the goal of finding Geometric or Euclidean Steiner tree is to connect given points by lines of minimum total distance in such a way that any two points may be interconnected by line segments either directly or via other points and line segments. The connecting line segments should not intersect each other.
%\end{definition}

Now we move on to defining Bilu-Linial stability for the Steiner tree
problem on metrics.

\begin{definition}[{\bf Bilu-Linial $\gamma$-stabile instances}]
Let $I = (G, w)$ be an instance of a metric Steiner tree problem and $\gamma > 1$. 
$I$ is $\gamma$-stable if for any function
$w': V \times V \rightarrow \mathbb{R}^{+}_{0}$ such that 
$\forall u,v \in V$,
$$
w_{uv} \le w'_{uv} \le \gamma w_{uv},
$$
the optimal Steiner tree $\mathrm{OPT}'$ under $w'$ is equal to the
optimal Steiner tree $\mathrm{OPT}$ under $w$.
\end{definition}

%\begin{definition}[{\bf Bilu-Linial Stability}]
%An instance is Bilu–Linial stable, also known as perturbation resilient, if the optimal solution does not change when we perturb the instance.
%\end{definition}

%\begin{definition}[$\gamma$-stable Steiner tree]
%consider an instance $I = (G, w)$ of a Steiner tree problem, an instance I is $\gamma$-stable for any $\gamma>1$ if every $\gamma$-perturbation of I has the same optimal solution as I.
%\end{definition}

We note that the perturbations can be
such that instances originally satisfying 
the metric or Euclidean properties
no longer have to satisfy these properties
after perturbation.
We also note that due to
the triangle inequality, no instances
have stability $2$ or greater in the metric setting.

\textbf{Notation:}
For a graph $G$, $w_{ab}^{G}$ is the weight of edge $ab$ in $G$. We abbreviate $w_{ab} = w_{ab}^{G}$ and $w'_{ab} = w^{G'}_{ab}$. Let $\mathrm{OPT} \subseteq E(G)$ denote the minimum weight Steiner tree of $G$, let $w(\mathrm{OPT}) = w^{G}(\mathrm{OPT})$ denote the weight of the Steiner tree.

\section{Structural properties in general metrics}

In this section, we work in the context of a general metric space, and we develop interesting restrictions on the types of problems with $\gamma$-stable solutions, for various values of $\gamma.$ 

The techniques of this section \emph{do not} give, in complete generality, an efficient algorithm for finding the optimal Steiner tree for any value of $\gamma$ less than $2,$ a problem we leave open. However, when more information about the metric space is available, one can use the structural results here to give restrictions on the arrangements of Steiner points which does yield a definitive solution. In particular,

\begin{enumerate}
\item In Section~\ref{sec:Euclidean}, we use Lemma~\ref{lem: Steiner degree} to give an algorithm for the Euclidean metric when $\gamma > \sqrt{2}$.

\item More generally, in the case that no two Steiner points are adjacent in the optimal solution, Lemma \ref{funfan} together with the other results of the section can be used to give an efficient and very simple algorithm to find the minimal weight Steiner tree. Other more general situations can be efficiently handled via only slightly more elaborate arguments - e.g. if one has a bound on the length of the longest path of Steiner points in the optimal solution. 
\end{enumerate}

\begin{lem} \label{lem: Steiner degree}
The degree of any Steiner point in the optimal solution is greater than $\frac{2}{2-\gamma}$. 
\end{lem}

\proof Consider a Steiner node $s$ in the optimal solution, that is connected to 
$(m\neq n) $ other points, $a_1,...,a_m$. Let $\overline{w} = \sum_{i = 1}^m\frac{w_{sa_i}}{m}$, and let $w_{sv_1}$ and $w_{sa_m}$ be such that $w_{sv_1} + w_{sa_m} \geq 2 \overline{w}$. Let $G'$ be obtained by perturbing each edge $sv_i$ by a factor of $\gamma$. Let 
\[
\mathrm{OPT}' := (\mathrm{OPT} \setminus \{sv_1,\ldots, sv_m\}) \cup \{v_1v_2,\ldots,v_{m-1}v_m\}.
\]

Clearly, $\mathrm{OPT}'$ is also a Steiner tree. Using the fact that $w_{a_ia_{i+2}} \leq w_{sa_i} + w_{s_{a_{i+1}}}$, we have

\begin{align*}
w'(\mathrm{OPT}') & \leq w'(\mathrm{OPT}) - \sum_{i=1}^{m}\gamma w_{sa_i} \\
&~~~~~+\sum_{i=1}^{m-1}\left(w_{sa_i} + w_{sa_{i+1}}\right) \\
& =  w'(\mathrm{OPT}) - \sum_{i=1}^{m}\gamma w_{sa_i} +\sum_{i=2}^{m-1}2w_{sa_i} \\
&~~~~~+ w_{sa_1} + w_{sa_m}
\end{align*}

Using the fact that $w'(\mathrm{OPT}') > w'(\mathrm{OPT})$, we have 

\[
\sum_{i=1}^{m}\gamma w_{sa_i} < \sum_{i=2}^{m-1}2w_{sa_i} + w_{sa_1} + w_{sa_m}
\]

or

\[
\gamma \cdot \overline{w} m < (2m-2)\overline{w}.
\]

Rearranging, we have 
\[
\frac{2}{2-\gamma} < m
\]

\qed

Now we state some additional structural properties of optimal Steiner trees in $\gamma$-stable instances. These are not used in Section~\ref{sec:Euclidean}. Nevertheless, we hope that they are of independent interest.

\begin{lem}
If $a,b \in V(\mathrm{OPT})$ are nearest neighbors in the graph, then the edge $ab$ is in the optimal solution. 
\end{lem}

\begin{lem}
\label{lem:basicprop}
Suppose $ab,bc \in \mathrm{OPT}$, then
\begin{enumerate}
\item $w_{ac} > \gamma \cdot \max \{ w_{ab} ,  w_{bc}\}$.

\item $\frac{2}{\gamma} \cdot w_{ac} > w_{ab} + w_{bc}$.

\item $(\gamma - 1) \cdot w_{ab} < w_{bc} $, $(\gamma - 1) \cdot w_{bc} < w_{ab}.$
\end{enumerate}

\end{lem}

\proof
\begin{enumerate}
\item Assume w.l.o.g. $w_{ab} \geq w_{bc}$. Suppose that $w_{ac} \leq \gamma \cdot\max \{w_{ab},w_{bc}\}$, let $G'$ be obtained by perturbing $ab$ by a factor of $\gamma$. Then $(\mathrm{OPT} \setminus \{ab\}) \cup \{ac\}$ is also a Steiner tree in $G'$ of weight $w'(\mathrm{OPT})$ contradicting stability. This completes the proof of $1.$
\item  The proof of $2.$ follows from $1.$ and the fact that $\max\{w_{ab}, w_{bc}\} \geq \frac{w_{ab} + w_{bc}}{2}$.
 
 \item Let $G'$ be obtained by perturbing $bc$ by a factor of $\gamma$. Then $\mathrm{OPT}' : = \mathrm{OPT} \setminus \{bc\} \cup \{ac\}$ is also a Steiner tree of weight 
\begin{equation}
\label{eqn:eqn1}
w'(\mathrm{OPT}') = w(\mathrm{OPT}) - w_{bc} + w_{ac} \leq w(\mathrm{OPT}) + w_{ab}.
\end{equation}
On the other hand, stability 
gives us that
\begin{equation}
\label{eqn:eqn2}
w'(\mathrm{OPT}') > w'(\mathrm{OPT}) = w(\mathrm{OPT}) + (\gamma - 1)w_{bc}.
\end{equation}
Putting~(\ref{eqn:eqn1}) and~(\ref{eqn:eqn2}) together gives us that $(\gamma - 1) \cdot w_{bc} \leq w_{ab}$.

Repeating the same argument but swapping $bc$ for $ab$ gives us $(\gamma - 1) \cdot w_{ab} \leq w_{bc}$.
\end{enumerate}
\qed

%\begin{proposition}
%\textcolor{red}{AP:check}
%Let $H$ be any connected subgraph of $\mathrm{OPT}$. Let $b \in V(H)$ and $a \in V(\mathrm{OPT}) \setminus V(H)$ satisfy: 
%\begin{enumerate} 
%\item $\forall b' \in H,$ $\gamma w_{ab} \leq w_{ab'}$. 
%\item $\forall b' \notin H,$ $\gamma w_{ab} \leq w_{ab'}+w_{b'b}$. 
%\end{enumerate}
%Then $ab \in \mathrm{OPT}$. 
%\end{proposition}

\begin{lem} \label{close} 
Let $H$ be a subgraph of $\mathrm{OPT}$ with at least one edge. Let $ab \in H.$  Fix any vertex $c \in V(\mathrm{OPT} )\setminus V( H)$ satisfying $w_{ca} \leq \gamma (\gamma -1 )  \cdot w_{ab}$; then we have $ca \in \mathrm{OPT}$. 
\end{lem} 

\proof
If $ca \notin \mathrm{OPT}$, then adding the edge $ac$ to $\mathrm{OPT}$ produces a cycle which includes edge $ac.$ Suppose that the cycle also includes $ab$. Let $G'$ be obtained by perturbing $ab$ by a factor of $\gamma$. Then $(\mathrm{OPT}' \setminus \{ab\}) \cup \{ac\}$ is a Steiner tree of weight at most $w'(\mathrm{OPT})$, contradicting stability.  

If the cycle does not include $ab$, it includes some edge other than $ac$ which has endpoint at $a$. This edge, call it $ad$, is in $\mathrm{OPT}$. By Lemma \ref{lem:basicprop}, $w_{ad} > (\gamma -1) w_{ba}$. Let $G'$ be obtained by perturbing $ad$ by a factor of $\gamma$. We have $w'_{ad} > \gamma (\gamma -1) w_{ba} \geq w_{ac}$. Then $(\mathrm{OPT} \setminus \{ad\}) \cup \{ac\}$ is a Steiner tree of weight less than $w(\mathrm{OPT})$, again contradicting stability.  
\qed

\begin{lem}
Let $\gamma > \frac{1+\sqrt{5}}{2}.$ Let $ab \in H$, a subgraph of $\mathrm{OPT}.$
Suppose that $c$ is a vertex with $w_{ca} \geq \gamma \cdot w_{ab}$, then $ca \notin \mathrm{OPT}$.
\end{lem}

\proof
Let $\gamma ' = \frac{w_{ca}}{w_{ab}}.$ Note that $\gamma' \geq \gamma$ is some real number larger than $\frac{1+\sqrt{5}}{2}.$ If $ac \in \mathrm{OPT}$, then by part $1.$ of Lemma~\ref{lem:basicprop}, we must have 
\[
\frac{w_{bc}}{w_{ac}} >\gamma.
\]

On the other hand, \begin{eqnarray*} 
\frac{w_{bc}}{w_{ac}} & \leq & \frac{w_{ab}+ w_{ac}}{w_{ac}} \\
& \leq & \frac{w_{ab}+ \gamma ' w_{ab}}{\gamma 'w_{ab}} \\
& \leq & \frac{1+\gamma'}{\gamma '}.
\end{eqnarray*}
We now have a contradiction as long as $\frac{1+\gamma'}{\gamma '} < \gamma $. The function $f(x) = \frac{1+x}{x}$ is decreasing for $x>0$ and $f(x) < x$ for any $x \geq \frac{1+\sqrt{5}}{2}$. So, we have that $\frac{1+\gamma '}{\gamma '} < \frac{1+\gamma }{\gamma } <\gamma $ as desired. 
\qed

\begin{proposition} \label{close2}
Let $H $ be a subgraph of $\mathrm{OPT}$ with at least one edge. Suppose that $ab \in H$ and suppose that $c \in V(\mathrm{OPT}) \setminus V(H)$ with $w_{bc} <\gamma (\gamma - 1) w_{ab}$. Then we must have $w_{bc} < \frac{w_{ab}}{\gamma -1}$ and $w_{ab} < \frac{w_{bc}}{\gamma-1} .$ 
\end{proposition}

\proof By Lemma \ref{close}, we must have that $bc \in \mathrm{OPT}$. Therefore, property $3.$ of Lemma~\ref{lem:basicprop} gives us the desired inequalities. 
\qed

When $\gamma (\gamma -1)^2 > 1$ Proposition \ref{close2} strengthens the bounds of Lemma \ref{close}. This holds, for instance, when $\gamma >1.755$. In this case, we obtain: 

\begin{proposition}
Assume that $\gamma (\gamma -1)^2 > 1$. Assume that $H$ is a subgraph of $\mathrm{OPT}$ with at least two vertices. Let $ab \in H.$  Fix any vertex $c \in V(\mathrm{OPT} )\setminus V( H)$. Then we have $w_{ca} < \frac{1}{\gamma -1} \cdot w_{ab}$ if and only if $ca \in \mathrm{OPT}$. 
\end{proposition}

\proof By Lemma \ref{close} and the assumption that $\gamma (\gamma -1) > \frac{1}{\gamma -1}$, we must have that $ac \in \mathrm{OPT}$. If $w_{ca} \geq \frac{1}{\gamma -1} \cdot w_{ab}$, we can not have edge $ac$ in $\mathrm{OPT}$ by Lemma \ref{lem:basicprop} part $3.$ 

\qed

Let $a, \bar b = (b_1, \ldots , b_m)$ be vertices (either terminal or Steiner points). We denote by $T(a, \bar b)$ the tree on vertex set $a,\bar b$ in which $a $ is connected to each element of $\bar b.$ Let the \emph{average weight of $T(a, \bar b)$} be 

\[
\frac{\sum_{i=1}^m w_{ab_i}}{m-1}.
\]

Suppose that $H$ is a subgraph of $\mathrm{OPT}$. We call $T(a,\bar b)$ a \emph{terminal component fan} relative to $H$ if $a$ is a Steiner point and $\bar b$ are all terminals or vertices in distinct connected components of $H$ each with at least two vertices. We call the collection of components of $H$ together with the terminals not in $H$ the \emph{terminal components of $H$.} 

\begin{lem} \label{funfan} 
Let $\gamma > 1.755$ and suppose that $H$ is a subgraph of $\mathrm{OPT}$ and in the optimal solution, no two Steiner points are adjacent. Suppose that $T(a,\bar b)$ with $\bar b = (b_1, \ldots , b_m)$ is a terminal component fan such that: 
\begin{itemize} 
\item the average weight of $T(a,\bar b)$ is less than all edges not in $H$ which connect two terminal components of $H$, 
\item the average weight of $T(a,\bar b)$ is minimal among all terminal component fans,
\item the edges of $T(a, \bar b)$ are all within a factor of $\frac{1}{\gamma - 1}$ of each other. 
\end{itemize} 
Then $T(a, \bar b)$ is a subgraph of $\mathrm{OPT}$. 
\end{lem}

\proof
Suppose that the fan $T(a, \bar b)$ is not in $\mathrm{OPT}.$ Then for some subset of the edges of $T(a,\bar b)$ are not in $\mathrm{OPT}$ - the components of $H$ that contain each $b_i$ are connected. Specifically, if there are $k < m$ edges of $T(a, \bar b)$ which are not in $\mathrm{OPT}$, then there are at least $k$ edges of $\mathrm{OPT} \setminus H$ such that in $\mathrm{OPT} \cup T(a, \bar b)$ we may remove these $k$ edges and still have a Steiner tree.\footnote{In the case that $k=m$, there may be only $m-1$ such edges, as $a$ may not be in $\mathrm{OPT}$, but the argument works identically in that case.} Moreover, since no two Steiner points are adjacent, these edges are either

%, we can find $k$ such edges which are 

\begin{itemize}
    \item terminal to terminal edges, or
    \item part of a terminal component fan. 
\end{itemize}

In the first case, the terminal to terminal edges have weight at least $\frac{\sum_{i=1}^m w_{ab_i}}{m}.$ In this case perturb this edge by a factor of $\gamma$, and swap it with one edge of the terminal component fan $T(a, \bar b)$. Since the edges of $T$ are within a factor of $\frac{1}{\gamma - 1}$ of each other and their average weight is $\frac{\sum_{i=1}^m w_{ab_i}}{m}$, this swap decreases of the weight of the resulting Steiner tree after the perturbation. 

Similarly in the case that one of the $k$ edges is in another terminal component fan, $T_1$, the average weight of edges in that fan is at least $\frac{\sum_{i=1}^m w_{ab_i}}{m}$, and applying part $3.$ of Lemma~\ref{lem:basicprop}, the minimal weight edge in $T_1$ is at least $(\gamma -1) \cdot \frac{\sum_{i=1}^m w_{ab_i}}{m}.$ Now, perturb such an edge by a factor of $\gamma$ to make the weight at least $\gamma \cdot (\gamma -1) \cdot \frac{\sum_{i=1}^m w_{ab_i}}{m},$ which is larger than the weight of the largest weight edge of $T(v, \bar w)$, which is a most $\frac{1}{\gamma-1} \cdot \frac{\sum_{i=1}^m w_{ab_i}}{m}$ because $\gamma > 1.755.$ 

Performing any of these $k$ swaps yields a lower weight Steiner tree than $\mathrm{OPT}$ under the above perturbations, contradicting $\gamma$-stability. 
\qed 

\iffalse
\begin{remark}
The techniques of this section \emph{do not} give, in complete generality, an efficient algorithm for finding the optimal Steiner tree for any value of $\gamma$ less than $2,$ a problem we leave open. However, when more information about the metric space is available, one can use the structural results here to give restrictions on the arrangements of Steiner points which does yield a definitive solution. For instance, in the next section, we do this for Euclidean space. 

More generally, in the case that no two Steiner points are adjacent in the optimal solution, Lemma \ref{funfan} together with the earlier results of the section can be used to give an efficient and very simple algorithm to find the minimal weight Steiner tree. Other more general situations can be efficiently handled via only slightly more elaborate arguments - e.g. if one has a bound on the length of the longest path of Steiner points in the optimal solution. 
\end{remark}
\fi

\section{Euclidean Steiner trees}\label{sec:Euclidean}

In this section, we consider the restriction of the Steiner tree problem
to the Euclidean metric. 

%{\bf [TODO: explain Steiner ratio conjecture (or theorem) and why it doesn't apply. give a "counterexample" in the form of a Euclidean instance stable above the 1/(Steiner ratio).]}

Under the assumption of $\gamma$ stability the min angle between two terminal points can be defined as a function of $\gamma$. 

\begin{definition}[angle]
%For a vertex $u$ in the optimal Steiner tree and two of its neighbors $v_1, v_2$
Let $a_1,a_2,b$ be points on a Euclidean metric. Then
we call $\angle a_1ba_2$ the \emph{angle} between $a_1, a_2$.
\end{definition}

\begin{lem}\label{lem:theta&gamma}
For a $\gamma$-stable instance of a Euclidean Steiner tree, the angle between two terminal points with respect to their
common Steiner neighbor in the tree should be greater than $2\sin^{-1}(\gamma/2)$.
\end{lem} 

\proof Lets assume, for a $\gamma$-stable instances of Steiner tree, the angle between two terminal points $a_1$, and $a_2$ at a Steiner point $b$ is $\theta$.
 %This lowest amount of $\theta$ happens when the points form a Isosceles triangle where the length of the sides is a and the base is oat most $\gamma$a.
Without loss of generality, let $w_{a_1b} =: w \geq w_{a_2b}$ . Clearly $w_{a_1a_2} > \gamma w$, since otherwise, perturbing edge $a_1b$ by a factor of $\gamma$ allows one to replace $a_1b$ by $a_1a_2$ in a minimal Steiner tree, contradicting stability. Let us use $\alpha$ to denote the angle $\angle a_1a_2b$. Clearly, $\alpha \geq \pi/2 - \theta/2$. Thus by the sine rule, we have 
\[
    \frac{\gamma w}{\sin \theta} < \frac{w_{a_1a_2}}{\sin\theta}=\frac{w}{\sin\alpha} \leq \frac{w}{\sin(\pi/2-\theta/2)}.
\]
Rearranging, we have
\begin{align*}
   \gamma\ &<\  \frac{\sin\theta}{\sin(\pi /2 -\theta/2)} \\
   &=\ \frac{2\sin(\theta/2)\cos(\theta/2)}{\cos(\theta/2)}\\
   &=\ 2\sin(\theta/2)
\end{align*}
as desired. 
\qed

Thus we immediately get the following Corollary.

\begin{corollary}
For a $\gamma$-stable instances of Steiner tree,  if $\gamma > \sqrt{2}$ then the angle $\theta$ between two terminal points is $\theta > \pi/2$.
\end{corollary}

\begin{figure}
    \centering
\includegraphics[width=.25\textwidth]{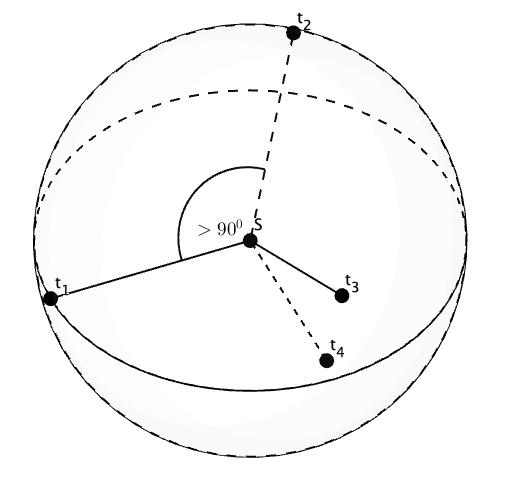}
    \caption{An example of points $t_1$, $t_2$, $t_3$, and $t_4$
    surrounding Steiner point $s$ at angles over $\theta > 90$ degrees. No more than $\frac{-1}{\mathrm{cos}{\theta}}$ can
    fit, independent of the dimension.}
    \label{fig:nosteiner}
\end{figure}

\begin{lem} 
If there are $N$ points in $\mathbb{R}^{d}$ such that the angle between every pair with respect to a point $u$ is at least $\theta > (\pi/2)$, then $N \leq \frac{-1}{\cos\theta}$.
\end{lem}

\proof
Let $\theta > \pi/2$ and let $v_1,\dots,v_N \in \mathbb{R}^{d} $ be unit vectors in $\mathbb{R}^{d}$ such that $\langle v_i,v_j \rangle \leq \cos\theta$. Consider the matrix $V$ whose columns are the $v_i$s. We know that by construction $V^TV$ is positive semi-definite. But if $N > \frac{-1}{\cos \theta}$, then the sum of every row is negative, which contradicts the positive semidefiniteness of $V^TV$, and so
it must be the case that $N \leq \frac{-1}{\cos\theta}$. 
\qed

\begin{corollary} \label{col:Steiner degree packing}
For $\gamma > \sqrt{2}$ the degree of a Steiner node in the optimal solution is at most $\frac{-2}{2 - \gamma^2}$.
\end{corollary}

\proof
From Lemma~\ref{lem:theta&gamma} we have
\[
\gamma< 2 \sin(\theta/2).\\
\]
So \[
\gamma^2 < 4 \sin^2(\theta/2)
\]
and so $\gamma^2/2 < 2\sin^2(\theta/2)$ or  $1- \gamma^2/2> 1- 2\sin^2(\theta/2)$.
Since $\cos( \theta) = 1- 2\sin^2(\theta/2)$, we have
\[
\cos (\theta)<1-\gamma^2/2
\]
or
\[
\theta > \cos^{-1} (1-\gamma^2/2).
\]
\qed

%\begin{lem}\label{lem: Steiner degree}
%The degree of any Steiner node in the optimal tree is at least d, $d = %\frac{3}{2-\gamma}.$ 
%\end{lem}
%\proof{Consider the Steiner node $s$ in the optimal solution, $s\in V_s$, that is connected to 
%$(m\neq n) $ other points, $v_1,...,v_m$ through the edge set $C$, $C = c_1, \ldots, c_m = c$, and all the $m$ vertices (points in metric space) are connected to their adjacent vertex through $d = \gamma c$. Then, by triangle inequality we can obtain the following bound of the degree of s, 

%\begin{align}
%    \sum_{i=1}^{m}\gamma c_i &\leq \sum_{i=1}^{m-1}c_i+c_{i+1}\\
%    &\leq c_1+\sum_{j=1}^{m-2} 2c_{j}+c_{m}\\
%     &\leq (2m-3)c 
%     \Rightarrow \frac{3}{2-\gamma} \leq m
%\end{align}
%}

\begin{corollary}
When $\gamma > 1.562$, 
%we have that $\frac{2}{\gamma^2-2} < \frac{3}{2 - \gamma}$, 
the optimal Steiner tree for a $\gamma$-stable instance
does not have Steiner nodes. 
%In other words, the optimal Steiner tree is just the Minimum Spanning Tree (MST) on terminals.
\end{corollary}

\proof
This happens when the min degree imposed by stability is larger than the max degree imposed by the packing bound.
By Lemmas~\ref{lem: Steiner degree} and \ref{col:Steiner degree packing} we have the following:\\
\[
\frac{2}{\gamma^2-2}\leq \frac{2}{2-\gamma}
\]
By solving the above equation for $\gamma$ we get $\gamma \geq \frac{\sqrt{17}-1}{2}$, which is bounded from above by $1.562$.
\qed

This geometric property implies
that for $1.562$-stable instances, Steiner points
will not be used in the optimal solution.  Hence,
an MST algorithm on just the terminal points will give
the answer in polynomial time.

%\section{Algorithm} \label{alg} 
%\subsection{Details}
%* Input \\
%* Output\\
%* Fan picking
%\subsection{Algorithm Analysis}

Finally, we point to the existence of Gilbert and Pollak's the Steiner ratio conjecture~\cite{GilbertP68}, which  states that in the Euclidean plane,
there always exists an MST within a cost of $2/\sqrt{3}$ of
the minimum Steiner tree,and the behavior of this ratio for higher dimensions is yet unknown.  Assuming this conjecture,
 in certain cases it may imply some limitations on the stability of Euclidean instances, especially in low dimensions, using the idea that even if the Steiner tree distances are ``blown up" by more than the Steiner ratio, one could instead use the MST instead and get a cheaper solution.  Unfortunately, because the MST may overlap with the Steiner tree, we cannot give a concrete statement.

\section{Using approximation algorithms to solve stable instances}

In this section we give a general argument about how strong approximation
algorithms for Steiner tree problems
give stability guarantees.  We note that it is known that an FPTAS for the Steiner tree would imply P=NP~\cite{ChlebikC08}, so there is no hope to use the result below in the general metric case.  But if at some future point an FPTAS for the Euclidean variant of the Steiner tree problem is developed (currently, only a PTAS is known to exist~\cite{Arora98}), then this would immediately imply the existence of polynomial-time algorithms for stable instances for any constant $\gamma >1$.

\begin{thm}
An FPTAS for the Steiner tree problem
gives a polynomial time algorithm for optimally solving any $\gamma$-stable
Steiner tree problem in time poly$(n, (\gamma-1)^{-1})$.  In particular,
this gives a polynomial-time algorithm for any constant $\gamma > 1$.
\end{thm}

\proof Assume we are given an FPTAS for the Steiner tree problem.
This means that we have an algorithm that runs in time
$\mathrm{poly}(n,1/\epsilon)$ on instances of size $n$ to
give $(1+\epsilon)$-approximations to the optimum Steiner tree.
Now consider a $\gamma$-stable instance for constant $\gamma > 1$.  
We run our FPTAS on that instance with $\epsilon = \frac{\gamma-1}{2n}$ to get an Steiner tree $S'$ with weight within
$\mathrm{OPT}(1+(\gamma-1)/2n)$.
We now claim that $S'$ must contain every edge in OPT whose weight is at least $\frac{\mathrm{OPT}}{n}$.  Suppose it doesn't -- then we could perturb
such an edge by $\gamma$ and increase the weight of the optimal solution
to $\mathrm{OPT}(1+(\gamma-1)/n)$ and 
$S'$ would become cheaper than $\mathrm{OPT}$, thereby violating $\gamma$-stability.

By the fractional pigeonhole principle, the most expensive edge of the FPTAS
satisfies the desired property above and is therefore in $\mathrm{OPT}$.  Hence, we can
contract this edge into a new vertex and get a new instance with $n-1$ vertices at $\gamma$-stability. 
We can continue this process, 
getting one new edge of the optimal in each iteration, until we have a constant-size problem that we can brute-force. 
\qed

We note that the above technique could be used to convert
event slightly weaker (than FPTAS) approximation algorithms to nontrivial stability guarantees.

\bibliography{paper}

\end{document}